\documentclass[a4paper, 10pt, oneside, onecolumn, notitlepage, final]{article}
\usepackage[text={160mm, 240mm}, left=25mm, vmarginratio=1:1]{geometry}
\usepackage{graphicx}
\usepackage{amssymb, amsfonts, amsmath, amsthm, mathtools, mathrsfs}
\usepackage{upgreek}
\usepackage[usenames, dvipsnames]{xcolor}
\usepackage[colorlinks, linkcolor=blue, anchorcolor=red, citecolor=blue]{hyperref}
\usepackage[backend=biber, sorting=none, maxbibnames=3, minbibnames=3, style=numeric-comp]{biblatex}
\usepackage{braket} 
\usepackage{authblk} 
\usepackage{orcidlink}
\usepackage{abstract}
\usepackage{array}
\usepackage{bigstrut}
\usepackage{titlesec}
\usepackage[T1]{fontenc}
\usepackage[utf8]{inputenc}
\usepackage{lmodern}
\usepackage[english]{babel}

\titleformat{\section}{\large\bfseries\centering}{\thesection}{0.5 em}{}
\titleformat{\subsection}{\normalsize\bfseries\centering}{\thesubsection}{0.5 em}{}

\begin{document}

\title{\bf Stochastic Thermodynamics of the Two-Dimensional Model of Transistors}

\author{\normalsize Jiayin Gu\orcidlink{0000-0002-9868-8186}\thanks{\texttt{gujiayin@njnu.edu.cn}}}
\affil{\normalsize School of Physics and Technology, Nanjing Normal University, Nanjing 210023, China}

\date{}
\maketitle

\begin{abstract}
We adopt a stochastic approach to study the charge transport in transistors. In this approach, the hole and electron densities are ruled by diffusion-reaction stochastic partial differential equations satisfying a local detailed balance condition. The electric field is supposed to be concentrated in very narrow regions around the two junctions and is also approximated to be static. In this way, not only are the laws of electricity, thermodynamics, and microreversibility consistent within this approach, but also the transistor can be easily modeled as a two-dimensional system. We perform the full counting statistics of the two coupled currents and the fluctuation theorem is shown to hold. Moreover, we show that the geometric shape of the transistor exerts great influence on the transport behavior. By modeling the transistor in two dimensions, the signal-amplification factor up to about $164$ can be achieved, which is comparable to the typical value of realistic transistors in the industry.
\end{abstract}

\section{Introduction}

\par Stochastic thermodynamics emerged in the past two decades provides a conceptual framework for describing a large
class of thermodynamic systems arbitrarily far from equilibrium~\cite{Seifert_RepProgPhys_2012, Peliti_2021, Shiraishi_2023}. For example, in recent years, we have seen its applications to study the electronic curcuits~\cite{Gao_PhysRevRes_2021, Freitas_PhysRevX_2021, Freitas_PhysRevE_2022}. In these studies, the elementary electronic devices are stochastically modeled with the coarse-grained state characterized by only a few degrees of freedom. In addition, stochastic thermodynamics can also be applied at a more detailed level. In 2009, Andrieux and Gaspard developed a stochastic approach to study transport of ions in conductive channels~\cite{Andrieux_JStatMech_2009}. In this approach, the electric field is not only generated externally, but also self-consistently incorporates the contributions of the local deviations from electroneutrality. In 2018 and 2019, this stochastic approach was extended to study the charge transport in $p$-$n$ junction diodes and bipolar $n$-$p$-$n$ junction transistors~\cite{Gu_PhysRevE_2018, Gu_PhysRevE_2019}.

\par Compared with the diode which can be easily modeled as a one-dimensional channel, the transistor has three ports and exhibits a much more irregular geometric shape. In order to simplify the calculation of the electric field from the charge distribution and boundary conditions according to the Poisson equation, we also modeled the transistor as a one-dimensional system in Ref.~\cite{Gu_PhysRevE_2019} (hereafter paper I). Our focus of paper I was on the fundamental issue of microreversibility in nonequilibrium statistical physics. It was shown that the two currents that are coupled together satisfy the fluctuation theorem~\cite{Evans_PhysRevLett_1993, Gallavotti_PhysRevLett_1996, Kurchan_JPhysA_1998, Lebowitz_JStatPhys_1999}. In addition, the statistical cumulants of the currents and their response coefficients are shown to be correlated, satisfying the Onsager reciprocal relations and their generalizations to nonlinear transport properties~\cite{Andrieux_JChemPhys_2004, Andrieux_JStatMech_2007, Gaspard_NewJPhys_2013, Barbier_JPhysA_2018}. In the technological aspect of the transistor, the coupling between two currents allows one current to be used to manipulate the other that is much larger -- the so-called signal-amplification effect~\cite{Shockley_PhysRev_1951, Brennan_2005, Colinge_2005, Neamen_2003}. In paper I, however, we only obtained a signal-amplification factor of $4.278$, which is far below the typical value of industrial transistors ranging between 50 and 300. With some detailed analysis, we realize that this is due to oversimplification in the modeling of the transistor as a one-dimensional system. It is expected that, if the transistor was modeled in two dimensions, much higher values of the signal-amplification factor would be achieved.

\begin{figure}
\centering
\begin{minipage}[t]{0.48\hsize}
\resizebox{1.0\hsize}{!}{\includegraphics{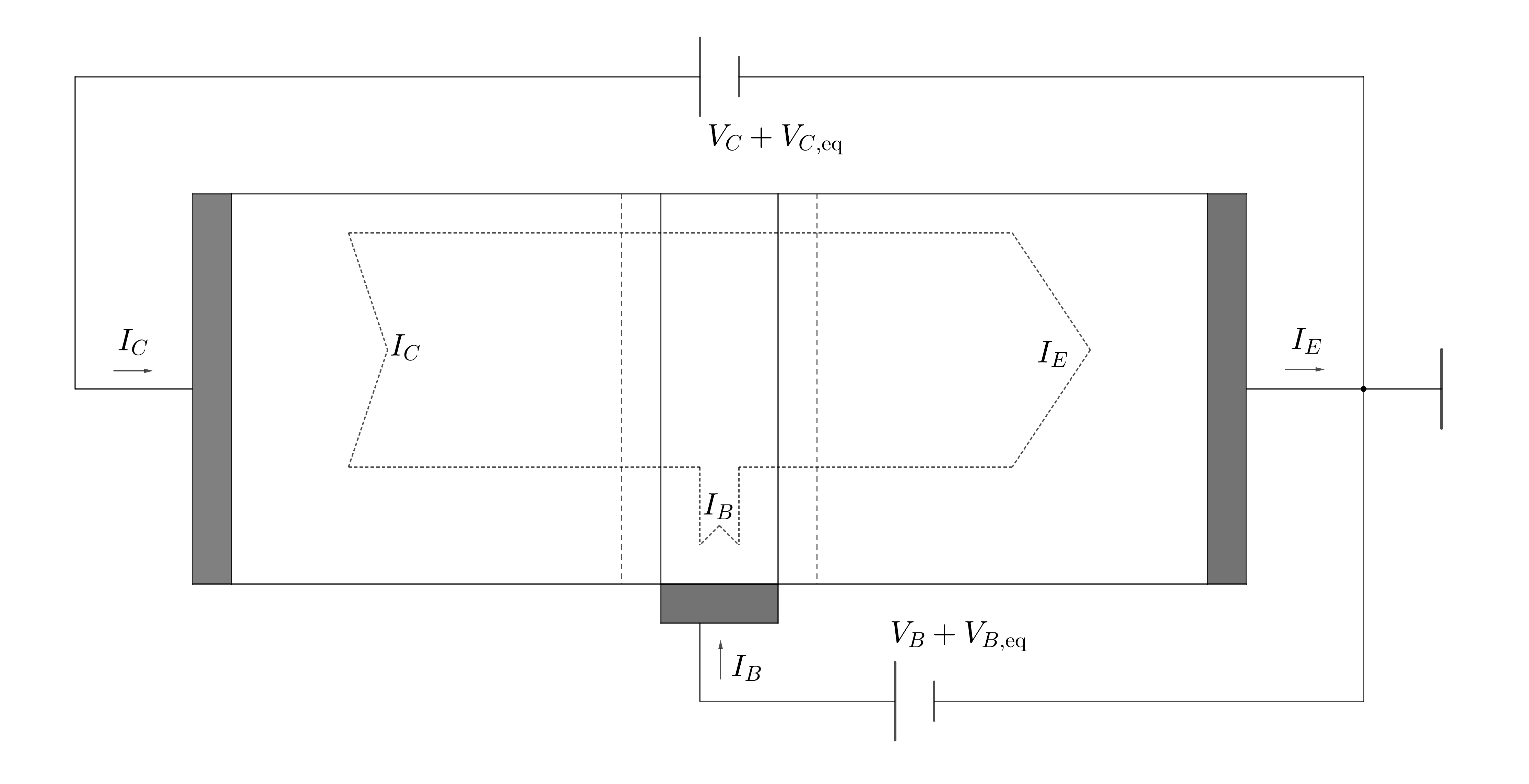}}
\end{minipage}
\begin{minipage}[t]{0.50\hsize}
\resizebox{1.0\hsize}{!}{\includegraphics{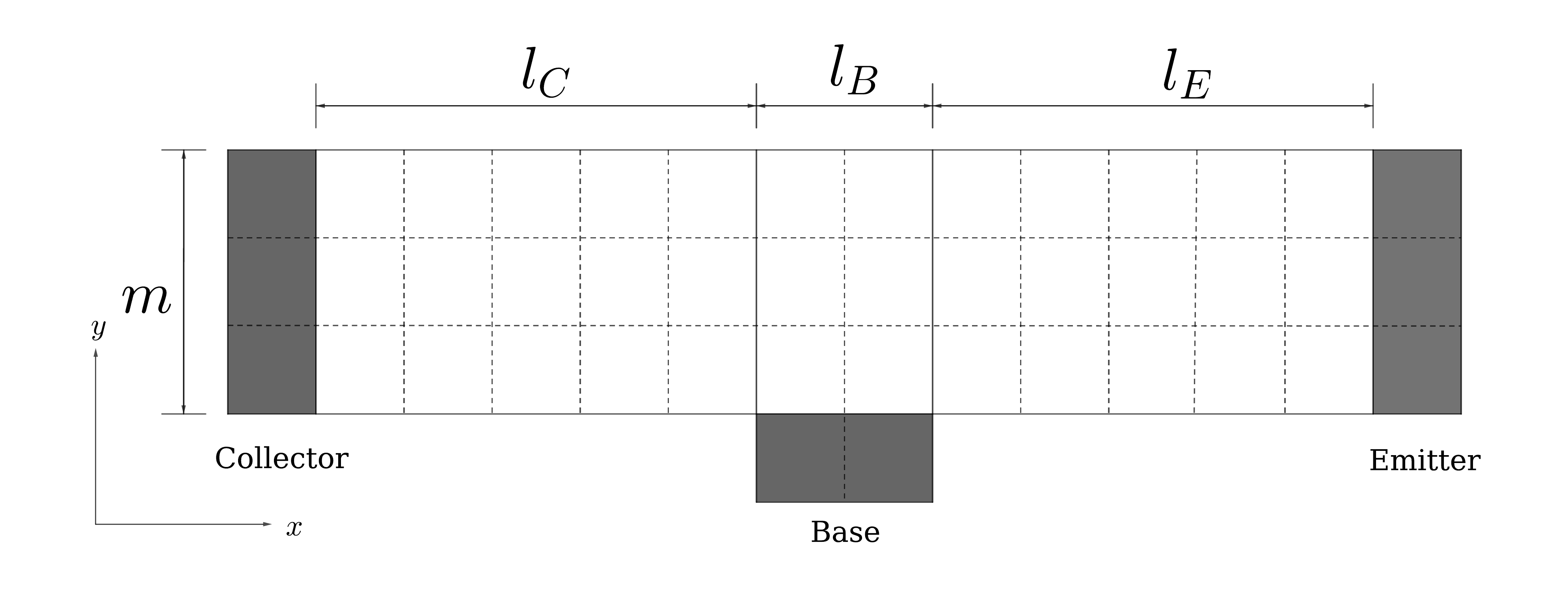}}
\end{minipage}
\caption{Schematic diagrams of bipolar $n$-$p$-$n$ junction transistors. In the left panel, it shows how the transistor is connected in a circuit and how currents flow inside it. In the right panel, the dimensions of the transistor are marked and how the system is spatially discretized into cells (delimited by dash lines) is shown. For illustrative purpose, this discretization scheme corresponds to Set I in Table~\ref{tab_values_2}. The terminals are in contact with three reservoirs (shaded regions) fixing the densities of holes and electrons. These reservoirs are here called \textit{Collector}, \textit{Base}, and \textit{Emitter}, respectively.}
\label{fig_transistor}
\end{figure}

\par The main purpose of this paper is to build a two-dimensional model of the transistor. The difficulty encountered is how to calculate the electric field according to charge distribution and boundary conditions which are not well defined. However, we observe that the electric field is actually concentrated on very narrow regions around two junctions in realistic transistors. This can be also seen from Fig. 2 in paper I that the profile of electric potential exhibits a steplike shape across the transistor. The sudden changes in the potential profile take place at junctions and the potential level of the steps are determined by the boundary conditions, i.e., the potentials of the three ports. This steplike shape of the profile is achieved when the concentration of majority charge carriers is overwhelmingly larger than that of minority charge carriers, as is required in realistic transistors. So, in modeling the transistor in two dimensions, we approximately fix the electric field and suppose that the fluctuation in the charge distributions has negligible influence on the electric field. In this way, we do not need to adaptively calculate the electric field according to the fluctuating charge distributions.

\par The vehicles of our study are as follows. In Sec.~\ref{sec_stochastic_description}, we present the stochastic description of the bipolar $n$-$p$-$n$ junction transistor, with a particular emphasis on the spatial configuration of the transistor as a two-dimensional system. The numerical method for simulating the transistor is presented. The transistor is spatially discretized and a master equation describing the stochastic evolution of the system state is established. In Sec.~\ref{sec_fluctuation}, we perform the full counting statistics of the charge transport. By approximating the charge transport with a coarse-grained model, the driving forces or affinities can be calculated from the statistics and their values are found in agreement with their theoretical expectations. In this way, the fluctuation theorem is indirectly tested. In Sec.~\ref{sec_functionality}, the functionality of the transistor is studied. Under proper working conditions, the signal-amplifying effect is realized and, moreover, its factor is shown to be influenced greatly by the geometric shape of the transistor. The conclusions and perspectives are drawn in Sec.~\ref{sec_conclusion}.

\section{Stochastic Description of Transistors}\label{sec_stochastic_description}

\subsection{Bipolar $n$-$p$-$n$ Junction Transistors}

\par As its name suggests, the bipolar $n$-$p$-$n$ junction transistor is composed of three semiconductors connected in series, with a thin $p$-type semiconductor sandwiched in between two $n$-type semiconductors, as shown in Fig.~\ref{fig_transistor}. Accordingly, two junctions are formed. The negative-charged acceptors and positive-charged donors are the impurities respectively doped in $p$-type and $n$-type semiconductors. They are anchored with uniform densities $a$, $d$ in their respective semiconductor. The other two kinds of charge carriers are mobile and they are positive-charged holes $h^+$ and negative-charged electrons $e^-$. The balance of charge requires that the number of holes is more than that of electrons in $p$-type semiconductors and there are more electrons than holes in $n$-type semiconductors. These mobile charge carriers diffuse across the transistor with the intensity characterized by the coefficient $D$~\footnote{For simplicity, both the diffusion coefficients for holes and electrons are assumed homogeneous across the transistor and equal here, i.e., $D_p({\bf r})=D_n({\bf r})=D$.}. The mobility of these charge carriers is related with their diffusion coefficient through the Einstein relation $\mu=\beta D$, where $\beta\equiv 1/(k_{\rm B}T)$ is the inverse temperature, $k_{\rm B}$ the Boltzmann constant, and $T$ the temperature. Besides, hole-electron pairs are generated and recombined according to the reactions
\begin{align}
\emptyset\xrightleftharpoons[k_-]{k_+} h^++e^- \text{,} \label{eq_reaction}
\end{align}
where $k_+$ and $k_-$ are the generation and recombination rate constants, respectively.

\par As shown in the right panel of Fig.~\ref{fig_transistor}, the width of the transistor is denoted by $m$ and the lengths of each part by $\{l_C,\,l_B,\,l_E\}$. These four parameters suffice to define the geometric shape of the transistor in two dimensions. Three terminals are in contact with three reservoirs called \textit{Collector}, \textit{Base}, and \textit{Emitter} in transistor terminology. The hole and electron densities as well as the electric potentials have fixed boundary values at contacts with the three reservoirs. They are respectively given by $p_C$, $n_C$, $\Phi_C$ at the \textit{Collector}, $p_B$, $n_B$, $\Phi_B$ at the \textit{Base}, and $p_E$, $n_E$, $\Phi_E$ at the \textit{Emitter}. A semiconducting material with no doped impurities is called intrinsic semiconductor. In this case, the hole and electron densities are equal and they are briefly denoted as the intrinsic density $\nu$. The whole transistor is fabricated by doping impurities into an intrinsic semiconductor of the density $\nu$. This induces the imbalance between the densities of holes and electrons. If the transistor is in equilibrium, we have $p_{\rm eq}n_{\rm eq}=\nu^2$ everywhere in the system. In this case, there is no flow of charge carriers and the detailed balance in the reaction~(\ref{eq_reaction}) requires that $k_+=k_-p_{\rm eq}n_{\rm eq}=k_-\nu^2$, which leads to $\nu=\sqrt{k_+/k_-}$. Moreover, in equilibrium the hole and electron densities at position ${\bf r}$ are given by
\begin{align}
p_{\rm eq}({\bf r})\sim{\rm e}^{-\beta e\Phi_{\rm eq}({\bf r})} \hspace{0.3cm}\text{and}\hspace{0.3cm} n_{\rm eq}({\bf r})\sim{\rm e}^{+\beta e\Phi_{\rm eq}({\bf r})}
\end{align}
in terms of the electric potential across the transistor. Since the reservoirs at the boundaries are always in equilibrium, they satisfy the conditions
\begin{align}
p_Cn_C=p_Bn_B=p_En_E=\nu^2 \text{.}
\end{align}
For simplicity, we set $p_C=p_E$ and $n_C=n_E$. In addition, the following boundary conditions,
\begin{align}
n_C=p_C+d \text{,}\hspace{0.3cm} p_B=n_B+a \text{,}\hspace{0.3cm} n_E=p_E+d \text{,}
\end{align}
are also imposed.

This electric potential, whether or not in equilibrium, is assumed to be determined completely by the boundary conditions at the contacts with the three reservoirs. Specifically, the electric potential across the left $n$-type semiconductor (respectively, middle $p$-type semiconductor and right $n$-type semiconductor) is uniform and given by $\Phi_C$ (respectively, $\Phi_B$, $\Phi_E$). This is the main difference from the case in paper I where the electric potential across the transistor is jointly determined by the Poisson equation together with the boundary conditions. Because charge carriers are inhomogeneously distributed across the transistor, Nernst potentials are produced,
\begin{align}
& V_{C,{\rm eq}}=\left(\Phi_C-\Phi_E\right)_{\rm eq}=\frac{1}{\beta e}\ln\frac{p_E}{p_C}=\frac{1}{\beta e}\ln\frac{n_C}{n_E} \text{,} \\
& V_{B,{\rm eq}}=\left(\Phi_B-\Phi_E\right)_{\rm eq}=\frac{1}{\beta e}\ln\frac{p_E}{p_B}=\frac{1}{\beta e}\ln\frac{n_B}{n_E} \text{.}
\end{align}
Here, $V_{C,{\rm eq}}=0$ according to the previous conditions $p_C=p_E$ and $n_C=n_E$ for simplicity purposes. When the transistor is driven out of equilibrium, the applied voltages can be defined with respect to the Nernst potentials
\begin{align}
& V_C=\Phi_C-\Phi_E-V_{C,{\rm eq}} \text{,} \label{eq_V_C} \\
& V_B=\Phi_B-\Phi_E-V_{B,{\rm eq}} \text{,} \label{eq_V_B}
\end{align}
and in this case currents are induced across the transistor. When $V_C-V_B<0$, the \textit{Collector}-\textit{Base} junction is said to be forward biased and when $V_C-V_B>0$ it is reverse biased. Similarly, the \textit{Emitter}-\textit{Base} junction is forward-biased when $V_B>0$, and reverse-biased when $V_B<0$. If no bias is applied, i.e., $V_C=V_B=0$, both junctions are in equilibrium and there is no charge ﬂow. A transistor with a forward-biased \textit{Emitter}-\textit{Base} junction and a reverse-biased \textit{Collector}-\textit{Base} junction is said to be operated in the forward active mode.

\subsection{Stochastic Diffusion-Reaction Equations}

\par The physical processes taking place inside the transistor are the drift and diffusion of holes and electrons, and reaction between them is described by Eq.~(\ref{eq_reaction}). The thermal agitation inside the transistor generates incessant errtic motion for the holes and electrons. This in turn causes local fluctuations in the currents and reaction rates. Stochastic approach is adopted to describe these fluctuations. In this approach, the Gaussian white noise fields are introduced in the diffusion-reaction equations for the hole and electron densities. Consequently, we can write down the following stochastic diffusion-reaction equations:
\begin{align}
& {\bf j}_p=+\mu ep\boldsymbol{\cal E}-D{\boldsymbol\nabla}p+\delta{\bf j}_p \text{,} \label{eq_diff_1} \\
& {\bf j}_n=-\mu en\boldsymbol{\cal E}-D{\boldsymbol\nabla}n+\delta{\bf j}_n \text{,} \label{eq_diff_2} \\
& \partial_tp+{\boldsymbol\nabla}\cdot{\bf j}_p=k_+-k_-pn+\delta\sigma \text{,} \label{eq_diff_3} \\
& \partial_tn+{\boldsymbol\nabla}\cdot{\bf j}_n=k_+-k_-pn+\delta\sigma \text{.} \label{eq_diff_4}
\end{align}
The electric field $\boldsymbol{\cal E}$ inducing the drift of holes and electrons is fixed and concentrates in very narrow regions around two junctions. The terms $-D{\boldsymbol\nabla}p$ and $-D{\boldsymbol\nabla}n$ are the currents due to diffusion, and the term $k_+-k_-pn$ denotes source and drain due to the generation and recombination of hole-electron pairs in the reaction. The terms $\delta{\bf j}_p$, $\delta{\bf j}_n$, and $\delta\sigma$ are Gaussian white noise fields associated with the hole diffusion, electron diffusion, and reaction, respectively. They are characterized by
\begin{align}
& \langle \delta{\bf j}_n({\bf r},t) \rangle = \langle \delta{\bf j}_p({\bf r},t) \rangle = 0 \label{av_j}\text{,} \\
& \langle \delta{\bf j}_n({\bf r},t)\otimes \delta{\bf j}_n({\bf r}',t') \rangle = \Gamma_{nn}({\bf r},t) \, \delta^3({\bf r}-{\bf r'}) \, \delta(t-t') \, {\boldsymbol{\mathsf 1}} \text{,} \\
& \langle \delta{\bf j}_p({\bf r},t)\otimes \delta{\bf j}_p({\bf r}',t') \rangle = \Gamma_{pp}({\bf r},t) \, \delta^3({\bf r}-{\bf r'}) \, \delta(t-t') \, {\boldsymbol{\mathsf 1}} \text{,} \\
& \langle \delta{\bf j}_n({\bf r},t)\otimes \delta{\bf j}_p({\bf r}',t') \rangle = 0 \text{,} \\
& \langle\delta\sigma({\bf r},t)\rangle = 0 \text{,} \label{av_s}\\
& \langle\delta\sigma({\bf r},t)\,\delta\sigma({\bf r'},t')\rangle = \Gamma_{\sigma\sigma}({\bf r},t) \, \delta^3({\bf r}-{\bf r'}) \, \delta(t-t') \text{,}  \\
& \langle \delta\sigma({\bf r},t)\, \delta{\bf j}_n({\bf r}',t') \rangle = \langle \delta\sigma({\bf r},t)\, \delta{\bf j}_p({\bf r}',t') \rangle = 0 \text{,}
\end{align}
where ${\boldsymbol{\mathsf 1}}$ is the $3\times 3$ identity matrix and
\begin{align}
& \Gamma_{pp}({\bf r},t)\equiv 2Dp({\bf r},t) \text{,} \\
& \Gamma_{nn}({\bf r},t)\equiv 2Dn({\bf r},t) \text{,} \\
& \Gamma_{\sigma\sigma}({\bf r},t)\equiv k_++k_-p({\bf r},t)n({\bf r},t)
\end{align}
are the noise spectral densities. The advantage of this approach is that the usual phenomenological parameters ($\mu$, $D$, $k_+$, and $k_-$) suffice for the stochastic description.

\subsection{Numerical Method for Simulating the Transistor}

\par For numerical simulation, the transistor is spatially discretized into cells, each with width $\Delta x$ in $x$-direction, width $\Delta y$ in $y$-direction, volume $\Omega$, and containing some numbers of holes and electrons. To indicate the position, we associate with each cell two indices $i$, $j$, where $i$ takes values $1,\,2,\cdots,L\equiv(l_C+l_B+l_E)/\Delta x$ for cells from left to right and $j$ takes values $1,\,2,\cdots,M\equiv m/\Delta y$ for cells from bottom to top. In this discretization scheme, as shown in the right panel of Fig.~\ref{fig_transistor}, the \textit{Collector} is modeled as $M$ cells, each with fixed $\bar{P}_C\equiv p_C\Omega$ holes and $\bar{N}_C\equiv n_C\Omega$ electrons. Likewise, the \textit{Base} is modeled as $l_B/\Delta x$ cells, each with fixed $\bar{P}_B\equiv p_B\Omega$ holes and $\bar{N}_B\equiv n_B\Omega$ electrons, and the \textit{Emitter} as $M$ cells, each with fixed $\bar{P}_E\equiv p_E\Omega$ holes and $\bar{N}_E\equiv n_E\Omega$ electrons. At the mesoscopic level of description, the transistor state is fully characterized by hole numbers ${\bf P}\equiv\{P_{i,j}\}$ and electron numbers ${\bf N}\equiv\{N_{i,j}\}$. Then, a Markov jump process in accord with Eqs.~(\ref{eq_diff_1})-(\ref{eq_diff_4}) can be associated to fully describe the stochastic evolution of the transistor state. The probability that the transistor is in the state $\{{\bf P},{\bf N}\}$ at time $t$ is ruled by the master equation. See Appendix~\ref{app_Markov} for the equation with the detailed explanation of the transition rates. This equation can be simulated in trajectory level with the standard Gillespie algorithm~\cite{Gillespie_JComputPhys_1976}. The state of the transistor changes every time when one charge carrier jumps between cells or a reactive event takes place. This implies that the simulation can be very time consuming, especially when there are a lot of holes and electrons in the transistor. When the numbers of holes and electrons in individial cells are very large (typically greater than 100), which is indeed the case, the Markov jump process can be approximated by the Langevin stochastic process. This allows a much faster simulation. In the continuum limit, the stochastic diffusion-reaction equations (\ref{eq_diff_1})-(\ref{eq_diff_4}) can be recovered from the resulting Langevin stochastic process. A detailed account is presented in Appendix~\ref{app_Langevin}.

\begin{table*}
\caption{Parameter values for the physics of semiconductors.}
\begin{center}
\begin{tabular}{>{\centering\arraybackslash}m{8.0cm}||>{\centering\arraybackslash}m{3.0cm}}
\hline
\hline
inverse temperature & $\beta=1.0$ \bigstrut \\ \hline
elementary charge & $e=|e|=1.0$ \bigstrut \\ \hline
diffusion coefficient for charge carriers & $D=0.01$ \bigstrut \\ \hline
hole-electron pairs generation rate constant & $k_+=0.01$ \bigstrut \\ \hline
hole-electron pairs recombination rate constant & $k_-=0.01$ \bigstrut \\ \hline
\hline
\end{tabular}
\end{center}
\label{tab_values_1}
\end{table*}

\par In numerical simulation, the statistical averages of any observable quantity $X$ can be evaluated by the time averaging $\langle X\rangle=\lim_{T\to\infty}(1/T)\int_0^TX(t){\rm d}t$, which is equivalent by ergodicity to the ensemble average $\langle X\rangle=\sum_{{\bf P},{\bf N}}{\cal P}_{\rm st}({\bf P},{\bf N})$ over the stationary probability distribution ${\cal P}_{\rm st}$. The densities of holes and electrons at the position ${\bf r}$ are recovered in the continuum limit as $p({\bf r})=P_{i,j}/\Omega$ and $n({\bf r})=N_{i,j}/\Omega$. These densiteis together with other quantities can be rescaled into dimensionless ones. See Appendix~\ref{app_rescaling} for a detailed account. This facilitates the comparison between the studies numerically and experimentally.

\section{Fluctuation Theorem for Currents}\label{sec_fluctuation}

\subsection{Generalities}

\par When the voltages~(\ref{eq_V_C}) and~(\ref{eq_V_B}) are applied at the boundaries, the transistor is driven out of equilibrium and after some transient time it relaxes to a nonequilibrium steady state. The \textit{Emitter} is taken as the reference reservoir; then two electric currents flow across the contact with the \textit{Collector} and across the contact with the \textit{Base}, respectively. They are coupled and their joint probability distribution satisfies the multivariate fluctuation theorem~\cite{Gaspard_NewJPhys_2013}. These electric currents are induced due to the random motion of holes and electrons crossing the contacting sections between the transistor and the corresponding reservoirs. So, we can define the instantaneous electric currents as
\begin{align}
& {\cal I}_C(t)\equiv\sum_{n=-\infty}^{+\infty}q_n^{(C)}\delta\left(t-t_n^{(C)}\right) \text{,} \\
& {\cal I}_B(t)\equiv\sum_{n=-\infty}^{+\infty}q_n^{(B)}\delta\left(t-t_n^{(B)}\right) \text{,}
\end{align}
where $t_n^{(C)}$ (respectively $t_n^{(B)}$) are the random times of the crossing events and $q_n^{(C)}$ (respectively $q_n^{(B)}$) are the transferred charges equal to $\pm e$ depending on whether the carrier is a hole or an electron and if its motion is inward or outward of the transistor. Further, the accumulated charges in unit $e$ over the time interval $[0,\,t]$ can be defined as
\begin{align}
& Z_C(t)\equiv\frac{1}{e}\int_0^t{\cal I}_C(t'){\rm d}t' \text{,}  \\
& Z_B(t)\equiv\frac{1}{e}\int_0^t{\cal I}_B(t'){\rm d}t' \text{.}
\end{align}
The fluctuation theorem for the joint probability distribution of the random variables $Z_C$ and $Z_B$ at time $t$ reads
\begin{align}
\frac{{\cal P}(Z_C,Z_B,t)}{{\cal P}(-Z_C,-Z_B,t)}\simeq_{t\to\infty}\exp\left(A_CZ_C+A_BZ_B\right) \text{,} \label{eq_FT}
\end{align}
where $A_C$ and $A_B$ are the driving forces also called affinities here. They are given by
\begin{align}
& A_C^{(t)}=\beta eV_C \text{,} \label{eq_A_C_t} \\
& A_B^{(t)}=\beta eV_B \text{.} \label{eq_A_B_t}
\end{align}
Here, the superscript $(t)$ is necessarily used to denote that they are expected in theory. The fluctuation theorem~(\ref{eq_FT}) implies that when positive voltages are applied the probability of observing a positive amount of charge transfers is far larger than that of observing the same amount of negative charge transfers. We define the mean currents of unit charge as
\begin{align}
& J_C\equiv\lim_{t\to\infty}\frac{1}{t}\langle Z_C(t)\rangle \text{,} \label{eq_J_C} \\
& J_B\equiv\lim_{t\to\infty}\frac{1}{t}\langle Z_B(t)\rangle \text{,} \label{eq_J_B}
\end{align}
where $\langle\rangle$ denotes the average over an ensemble of trajectories. Then, the mean electric currents are given by $I_C\equiv eJ_C$ and $I_B\equiv eJ_B$. Accordingly, the entropy production rate in units of $k_{\rm B}$ is expressed as
\begin{align}
\frac{1}{k_{\rm B}}\frac{{\rm d}_{\rm i}S}{{\rm d}t}=A_CJ_C+A_BJ_B=\beta\left(V_CI_C+V_BI_B\right)\ge 0
\end{align}
in terms of the dissipated power. This entropy production rate is always positive, in accordance with the second law of thermodynamics.

\begin{table*}
\caption{Parameter values for specifying spatial dimensions and charge distributions of the discretized transistor. Three sets of values are listed for different purposes.}
\begin{center}
\begin{tabular}{>{\centering\arraybackslash}m{6.5cm}||>{\centering\arraybackslash}m{2.0cm}|>{\centering\arraybackslash}m{2.0cm}|>{\centering\arraybackslash}m{2.0cm}}
\hline
\hline
Meaning & Set I  & Set II & Set III \bigstrut \\ \hline
width of the system, $m$ & $0.3$ & $0.5$ & $1.0$ \bigstrut \\ \hline
length of the system, $\{l_C,l_B,l_E\}$ & $\{0.5,0.2,0.5\}$ & $\{1.0,0.1,1.0\}$ & $\{1.0,0.1,1.0\}$ \bigstrut \\ \hline
dimensions of each cell, $(\Delta x,\Delta y)$ & $\{0.1,0.1\}$ & $\{0.1,0.1\}$ & $\{0.1,0.1\}$ \bigstrut \\ \hline
volume of each cell, $\Omega$ & $1000$ & $10^9$ & $10^9$ \bigstrut \\ \hline
number of electrons in a \textit{Collector} cell, $\bar{N}_C$ & $10000$ & $10^{13}$ & $10^{13}$ \bigstrut \\ \hline
number of holes in a \textit{Collector} cell, $\bar{P}_C$ & $100$ & $10^5$ & $10^5$ \bigstrut \\ \hline
number of electrons in a \textit{Base} cell, $\bar{N}_B$ & $100$ & $10^8$ & $10^8$ \bigstrut \\ \hline
number of holes in a \textit{Base} cell, $\bar{P}_B$ & $10000$ & $10^{10}$ & $10^{10}$ \bigstrut \\ \hline
number of electrons in an \textit{Emitter} cell, $\bar{N}_E$ & $10000$ & $10^{13}$ & $10^{13}$ \bigstrut \\ \hline
number of holes in an \textit{Emitter} cell, $\bar{P}_E$ & $100$ & $10^5$ & $10^5$ \bigstrut \\ \hline
\hline
\end{tabular}
\end{center}
\label{tab_values_2}
\end{table*}

\begin{table*}
\caption{Mean currents and their diffusivities evaluated from the full counting statistics of charge transport in different cases of affinities. The parameter values listed in Table~\ref{tab_values_1} and Set I in Table~\ref{tab_values_2} are used. For each case, the statistical data were obtained in simulation with the time step ${\rm d}t=0.05$, total time $t=1.0\times 10^4$, and $2.0\times 10^4$ iterates. }
\begin{center}
\begin{tabular}{>{\centering\arraybackslash}m{1.0cm}|>{\centering\arraybackslash}m{1.5cm}|>{\centering\arraybackslash}m{1.5cm}|>{\centering\arraybackslash}m{1.5cm}|>{\centering\arraybackslash}m{1.5cm}|>{\centering\arraybackslash}m{1.5cm}|>{\centering\arraybackslash}m{1.5cm}|>{\centering\arraybackslash}m{1.5cm}}
\hline
\hline
case & $A_C^{({\rm t})}$  &  $A_B^{({\rm t})}$  &  $J_C$  &  $J_B$  &  $D_{CC}$  &  $D_{BB}$  &  $D_{CB}$ \bigstrut \\ \hline
(1) & 0.000  &  0.000  &  -0.0065  &  -0.0134  &  387.817   &   442.942   &   -225.078   \bigstrut \\ \hline
(2) & 0.500  &  0.200  &  136.314  &  -10.589  &  362.842   &   438.204   &   -201.745   \bigstrut \\ \hline
(3) & 1.000  &  0.500  &  256.838  &  43.063   &  371.145   &   463.234   &   -190.971   \bigstrut \\ \hline
(4) & 1.500  &  0.700  &  376.805  &  73.434   &  371.770   &   477.123   &   -172.988   \bigstrut \\ \hline
\hline
\end{tabular}
\end{center}
\label{tab_affinities}
\end{table*}

\subsection{Numerical Results}

\par The direct test of the fluctuation theorem~(\ref{eq_FT}) requires the comparison between ${\cal P}(Z_C,Z_B,t)$ and its symmetric version ${\cal P}(-Z_C,-Z_B,t)$. As time $t$ goes to infinity, it usually becomes extremely difficult or even impractical to accurately determine the probabilities from the counting statistics of $Z_C$ and $Z_B$ over the overlapped region of these two distributions, especially when the affinities are very large. To overcome this issue, we have developed a coarse-grained model in paper I, which can be used to test the fluctuation theorem in an indirect way. Specifically, we first numerically estimate the affinities from the mean currents~(\ref{eq_J_C}) and (\ref{eq_J_B}) together with their diffusivities defined by
\begin{align}
& D_{CC}\equiv\lim_{t\to\infty}\frac{1}{2t}\left[\langle Z_C(t)Z_C(t)\rangle-\langle Z_C(t)\rangle^2\right] \text{,} \label{eq_D_CC} \\
& D_{BB}\equiv\lim_{t\to\infty}\frac{1}{2t}\left[\langle Z_B(t)Z_B(t)\rangle-\langle Z_B(t)\rangle^2\right] \text{,} \label{eq_D_BB} \\
& D_{CB}\equiv\lim_{t\to\infty}\frac{1}{2t}\left[\langle Z_C(t)Z_B(t)\rangle-\langle Z_C(t)\rangle\langle Z_B(t)\rangle\right] \text{.} \label{eq_D_CB}
\end{align}
Then, we compare the numerically obtained affinities with their theoretical expectations~(\ref{eq_A_C_t}) and (\ref{eq_A_B_t}). If agreements are found, we conclude that the fluctuation theorem~(\ref{eq_FT}) is indirectly tested. In this coarse-grained model, the long-time behavior of the charge transport is simplified at the highest level of description by
\begin{align}
& \textit{Collector}\xrightleftharpoons[W_{EC}]{W_{CE}} \textit{Emitter} \text{,} \\
& \textit{Base}\xrightleftharpoons[W_{EB}]{W_{BE}} \textit{Emitter} \text{,} \\
& \textit{Collector}\xrightleftharpoons[W_{BC}]{W_{CB}} \textit{Base} \text{,}
\end{align}
where the charge carriers are supposed to jump between the three reservoirs directly with the global transition rates $\{W_{kl}\}_{k,l=C,B,E}$. It is required that this coarse-grained model has the same values for the mean currents and their diffusivities as those of realistic transistors. So, we have the following equations for the global transition rates:
\begin{align}
& W_{CE}-W_{EC}+W_{CB}-W_{BC}=J_C \text{,} \label{eq_nonlinear_1} \\
& W_{BE}-W_{EB}+W_{BC}-W_{CB}=J_B \text{,} \label{eq_nonlinear_2} \\
& W_{CE}+W_{EC}+W_{CB}+W_{BC}=2D_{CC} \text{,} \label{eq_nonlinear_3} \\
& W_{BE}+W_{EB}+W_{BC}+W_{CB}=2D_{BB} \text{,} \label{eq_nonlinear_4} \\
& W_{CB}+W_{BC}=-2D_{CB} \text{.} \label{eq_nonlinear_5}
\end{align}
The affinities between the reservoirs in this coarse-grained model are given by
\begin{align}
& A_C^{(n)}\equiv A_{CE}^{(n)}=\ln\frac{W_{CE}}{W_{EC}} \text{,} \label{eq_A_C_n} \\
& A_B^{(n)}\equiv A_{BE}^{(n)}=\ln\frac{W_{BE}}{W_{EB}} \text{,} \label{eq_A_B_n} \\
& A_{CB}^{(n)}=\ln\frac{W_{CB}}{W_{BC}} \text{,} \label{eq_A_CB_n}
\end{align}
where the superscript $(n)$ denotes that they are affinities obtained numerically. The natural condition for these affinities is that
\begin{align}
A_{CB}^{(n)}+A_{BE}^{(n)}=A_{CE}^{(n)} \text{,}
\end{align}
which leads to
\begin{align}
W_{CB}W_{BE}W_{EC}=W_{BC}W_{EB}W_{CE} \text{.} \label{eq_nonlinear_6}
\end{align}
The Equations~(\ref{eq_nonlinear_1})-(\ref{eq_nonlinear_5}) together with Eq.~(\ref{eq_nonlinear_6}) form a set of nonlinear equations for the global transition rates. The values of mean currents $\{J_C,\,J_B\}$ and their diffusivities $\{D_{CC},\,D_{BB},\,D_{CB}\}$ are determined from full counting statistics in the numerical simulation. This set of nonlinear equations can be solved with the Newton-Raphson method~\cite{Press_2007}. Once the values of global transition rates are obtained, the numerical affinities can be readily evaluated according to Eqs.~(\ref{eq_A_C_n})-(\ref{eq_A_CB_n}). The description with the coarse-grained model is valid in the near-equilibrium regime because in this regime, the coarse-grained model is compatible with the fluctuation-dissipation relation. The advantage of using the coarse-grained model lies in that only the first- and second-order cumulants are required. These cumulants are relatively easy to access in simulation.

\begin{figure}
\centering
\begin{minipage}[t]{0.6\hsize}
\resizebox{1.0\hsize}{!}{\includegraphics{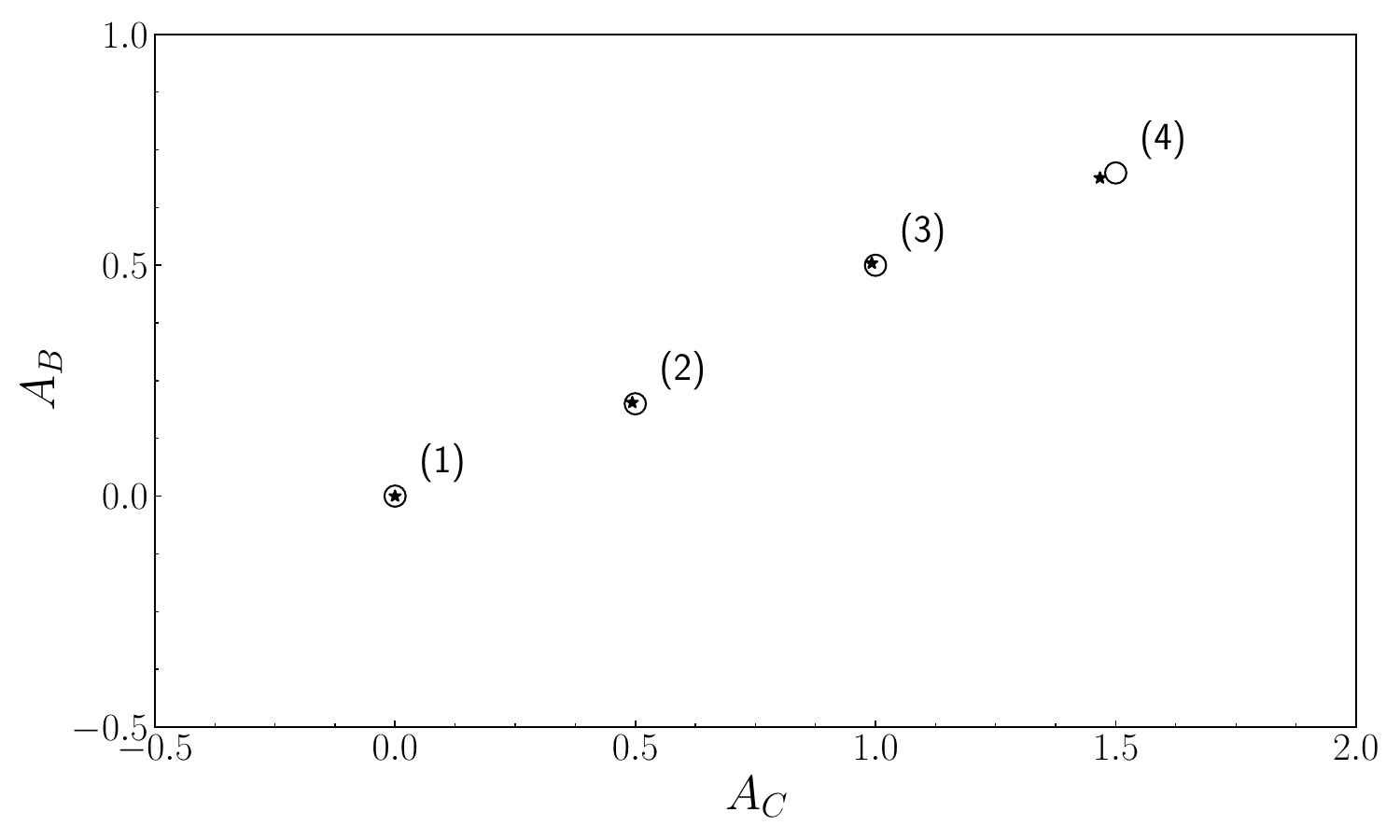}}
\end{minipage}
\caption{Comparison between numerical affinities calculated by Eqs.~(\ref{eq_A_C_n}) and (\ref{eq_A_B_n}) (marked with asterisks) and their theoretical expectations given by Eqs.~(\ref{eq_A_C_t}) and (\ref{eq_A_B_t}) (marked with circles).}
\label{fig_comparison}
\end{figure}

\par We carried out the numerical simulation with the parameter values listed in Table~\ref{tab_values_1} and Set I in Table~\ref{tab_values_2}. The results in different cases of theoretical affinities are listed in Table~\ref{tab_affinities}. The numerical affinities were calculated with the method above and their comparison between the corresponding theoretical expectations are drawn in Fig.~\ref{fig_comparison}. In this figure, the general agreement between these two kinds of affinities is found. Although we know that the fluctuation theorem should always hold for the currents, this test can be used to check whether the computer program for simulation is correctly coded. Actually, this is the primary motivation underlying this test.

\begin{figure}
\centering
\begin{minipage}[t]{0.95\hsize}
\resizebox{1.0\hsize}{!}{\includegraphics{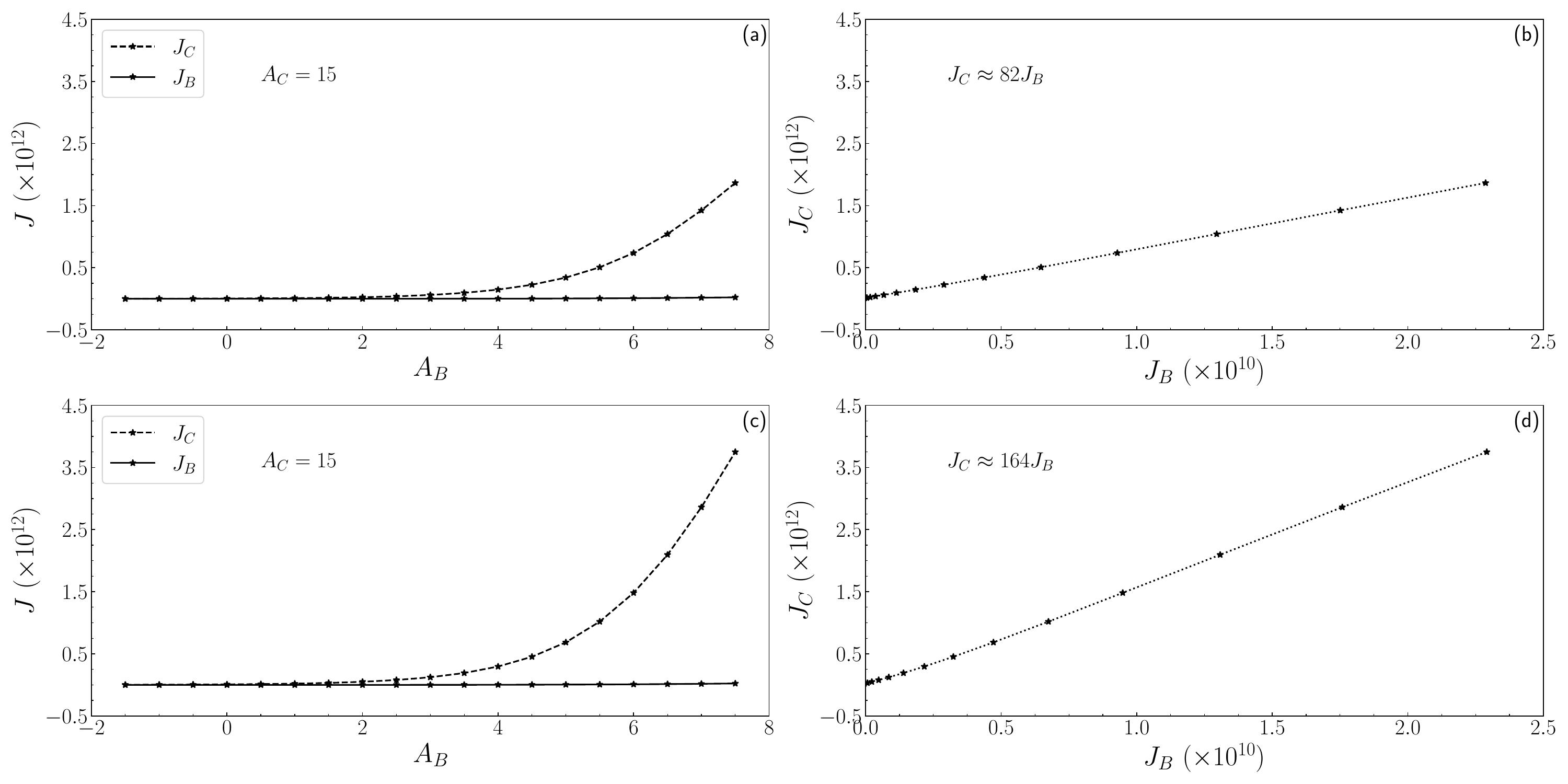}}
\end{minipage}
\caption{Diagrams showing the signal-amplifying effect. It shows in panels (a)-(c) how mean currents $I_C$ and $I_B$ vary as the affinity $A_B$ changes, with the other affinity fixed to the value $A_C=15$, and in panels (b)-(d) the corresponding relationships between the two mean currents. Approximated linear relationships are shown in the respective panels. The parameter values listed in Table~\ref{tab_values_1} and Set II in Table~\ref{tab_values_2} are used for panels (a) and (b). The parameter values listed in Table~\ref{tab_values_1} and Set III in Table~\ref{tab_values_2} are used for panels (c) and (d). In all panels, the lines join the numerical points depicted by asterisks. The simulations were carried out with the time step ${\rm d}t=0.001$ and total time $t=1.0\times 10^4$ for each data point.}
\label{fig_amplification}
\end{figure}

\section{The Functionality of Transistors}\label{sec_functionality}

\par Transistors are mainly used in semiconductor industry to amplify analog signals. To achieve this functionality, the transistor should be fabricated with the following conditions.
\begin{itemize}
\item[(1)] The doping impurities in \textit{Base} region should have a much lower density than those of the two neighboring regions,
\item[(2)] The concentration of the majority charge carriers in the \textit{Collector} region should be overwhelmingly larger than the concentration of minority charge carriers in the \textit{Base} region,
\item[(3)] The concentration of the majority charge carriers in the \textit{Emitter} region should be overwhelmingly larger than the concentration of minority charge carriers in the \textit{Base} region,
\item[(4)] The \textit{Base} region should be very thin so that the majority charge carriers in the \textit{Emitter} region can easily get swept to the \textit{Collector} region,
\item[(5)] The contacting section areas $\Sigma_C$ and $\Sigma_E$ should be larger than $\Sigma_B$.
\end{itemize}
The first three conditions are about the distributions of charge carriers and the last two about the geometric shape. The parameter values listed in Table~\ref{tab_values_1} together with Set II or Set III in Table~\ref{tab_values_2} are chosen to fulfill these conditions. It is necessary here to point out the implication of the first condition. The impurities doped in the intrinsic semiconductor cause the imbalance between the majority and minority charge carriers while still preserving $p_{\rm eq}n_{\rm eq}=\nu^2$. The much lower density of impurities in the \textit{Base} region implies that the extent of this imbalance is much milder than those of the other two regions. This is indeed the case in the parameter values listed in Set II and Set III in Table~\ref{tab_values_2}. By modeling the transistor in two dimensions, we have much more freedom in choosing the values of geometric parameters $\{m,\,l_C,\,l_B,\,l_E\}$. For the deliberately chosen values listed in Set II and Set III in Table~\ref{tab_values_2}, $l_B$ is very small and $w>l_B$. The last two conditions are satisfied. In addition, the following working conditions should also be satisfied.
\begin{itemize}
\item[(6)] The \textit{Collector}-\textit{Base} junction should be reverse biased,
\item[(7)] The \textit{Emitter}-\textit{Base} junction should be forward biased. In other words, the transistor should operate in the forward active mode, i.e., $A_C>A_B>0$.
\end{itemize}

\par As shown in the left panel of Fig.~\ref{fig_transistor}, there are two loops in the circuit. They have a common terminal, the \textit{Emitter}, which is often grounded. As such, this circuit configuration is said to be "common-\textit{Emitter}." The current $J_B$ in the \textit{Base}-\textit{Emitter} loop is regarded as an input and the current $J_C$ in the \textit{Collector}-\textit{Emitter} loop is correspondingly regarded as an output. These two currents are coupled so that it enables us to control the output current by changing the input current. This is called the transistor effect. The amplification factor is defined as the ratio between these two currents,
\begin{align}
\bar{\alpha}=\left(\frac{J_C}{J_B}\right)_{A_C} \text{.}
\end{align}
The differential amplification factor is defined as the ratio of the current variations induced by a slight change of $A_B$ while keeping $A_C$ fixed,
\begin{align}
\alpha=\left(\frac{\partial J_C}{\partial J_B}\right)_{A_C} \text{.}
\end{align}
Since the variations can be seen as signals, the latter is often called the signal-amplification factor. In ideal cases, these two amplification factors are approximately equal with each other, $\alpha\approx\bar{\alpha}$. So, the signal-amplification factor can be obtained directly from $J_B$ and $J_C$, which are much easier to measure.

\par The results are presented in Fig.~\ref{fig_amplification}. As $A_B$ increases with $A_C$ fixed, we observe that $J_B$ and $J_C$ both increase. Furthermore, we observe an approximately linear relation between these two currents. This is an expected relation from which we indeed have $\alpha\approx\bar{\alpha}$. For the purpose of contrast, two sets of results are provided in this figure. The top two panels correspond to the result obtained with the parameter values listed in Table~\ref{tab_values_1} and Set II in Table~\ref{tab_values_2}, while the bottom two panels correspond to the result with the parameter values listed in Table~\ref{tab_values_1} and Set III in Table~\ref{tab_values_2}. The only difference between Set III and Set II is that the transistor width $w$ is as twice large in Set III. The majority of the electrons diffused from the \textit{Emitter} region into the \textit{Base} region eventually drift into the \textit{Collector} region. Twice larger width implies that a twice as large amount of electrons get swept to the \textit{Collector} region. It is expected that the amplification factor is also twice as large for parameter values in Set III than that for parameter values in Set II. The numerical results show that this is indeed the case. From panels~(c) and (d) in Fig.~\ref{fig_amplification}, the amplification factor is
\begin{align}
\alpha\approx\bar{\alpha}\approx 164 \text{,} \label{eq_factor}
\end{align}
which is much larger than 4.278 that is obtained in paper I. This amplification factor~(\ref{eq_factor}) is comparable with those of realistic transistors in the industry.

\section{Conclusion and Perspectives}\label{sec_conclusion}

\par A stochastic approach has been adopted to study the charge transport in transistors. In this approach, we have approximated the electric field to be static in a stochastic approach to charge transport in transistors. In other words, there is no need to adaptively calculate the electric field once the charge distribution changes. This simplification does not sacrifice any main features of transistors, and more importantly, it has enabled us to easily model the transistor as a two-dimensional system. The electric field is supposed to be concentrated in very small regions around the two junctions of transistors. This is very reasonable especially for the case of realistic transistors where the concentration of majority charge carriers is overwhelmingly larger than that of minority charge carriers. In the stochastic approach, the charge carriers undergo processes of diffusion and reaction in the respective three regions of the transistor. When crossing the two junctions, the charge carriers undergo drift induced by the concentrated electric field and the local detailed balance condition is satisfied. The scheme is consistent with the laws of electricity, thermodynamics, and microreversibility.

\par The two-dimensional transistor system is discretized in space and a master equation has accordingly been introduced to describe the stochastic evolution of the system state. It has been shown that the fluctuation theorem holds for the two coupled currents in the transistor. In contrast with paper I, which mainly addresses the fundamental issue of microreversibility in nonequilibrium statistical physics, we have instead put more focus in this paper on the technological aspects. We have shown that the geometric shape of the transistor greatly influences the transport behavior. By modeling the transistor in two dimensions, we have realized a signal-amplification factor up to $164$. This value is comparable to that of realistic transistors. The stochastic approach supplied with suitable approximations finds a much broader application in its practical use.  The implication is that more electronic devices with technological interest are foreseeably to be studied in the future.

\section*{Acknowledgement}

\par This work was financially supported by startup funding from Nanjing Normal University.

\appendix

\section{Discretized Markov Jump Process}\label{app_Markov}

\par The probability distribution ${\cal P}({\bf P},{\bf N},t)$ to find the transistor in a certain state at time $t$ is ruled by the master equation
\begin{align}
\frac{{\rm d}{\cal P}}{{\rm d}t} = & \sum_{X=P,N}\sum_{j=1}^M\sum_{i=1}^{L-1}\left[\left({\rm e}^{+\partial_{X_{i,j}}}{\rm e}^{-\partial_{X_{{i+1},j}}}-1\right)W_{ij,x}^{(+X)}{\cal P}+\left({\rm e}^{-\partial_{X_{i,j}}}{\rm e}^{+\partial_{X_{{i+1},j}}}-1\right)W_{ij,x}^{(-X)}{\cal P}\right] \nonumber \\
& + \sum_{X=P,N}\sum_{i=1}^L\sum_{j=1}^{M-1}\left[\left({\rm e}^{+\partial_{X_{i,j}}}{\rm e}^{-\partial_{X_{i,{j+1}}}}-1\right)W_{ij,y}^{(+X)}{\cal P}+\left({\rm e}^{-\partial_{X_{i,j}}}{\rm e}^{+\partial_{X_{i,{j+1}}}}-1\right)W_{ij,y}^{(-X)}{\cal P}\right] \nonumber \\
& + \sum_{i=1}^L\sum_{j=1}^M\left[\left({\rm e}^{-\partial_{P_{i,j}}}{\rm e}^{-\partial_{N_{i,j}}}-1 \right)W_{ij}^{(+)}{\cal P}+\left({\rm e}^{+\partial_{P_{i,j}}}{\rm e}^{+\partial_{N_{i,j}}}-1 \right)W_{ij}^{(-)}{\cal P}\right] \nonumber \\
& + \sum_{X=P,N}\sum_{j=1}^{M}\left[W_{Cj,x}^{(+X)}\left({\rm e}^{-\partial_{X_{1,j}}}-1\right){\cal P}+\left({\rm e}^{+\partial_{X_{1,j}}}-1\right)W_{1j,x}^{(-X)}{\cal P}\right] \nonumber \\
& + \sum_{X=P,N}\sum_{j=1}^{M}\left[\left({\rm e}^{+\partial_{X_{L,j}}}-1\right) W_{Lj,x}^{(+X)} {\cal P}+W_{Ej,x}^{(-X)}\left({\rm e}^{-\partial_{X_{L,j}}}-1\right){\cal P}\right] \nonumber \\
& + \sum_{X=P,N}\sum_{i=l_C/\Delta x+1}^{(l_C+l_B)/\Delta x}\left[W_{iB,y}^{(+X)}\left({\rm e}^{-\partial_{X_{i,1}}}-1\right){\cal P}+\left({\rm e}^{+\partial_{X_{i,1}}}-1\right)W_{i1,y}^{(-X)}{\cal P}\right] \text{,} \label{eq_master_equation}
\end{align}
where the first and second lines respectively represent the contributions from the jump processes of charge carriers in the transistor in $x$-direction and $y$-direction, the third line from the contribution of reactions, and the remaining three lines respectively from contributions of the jump processes of holes and electrons between the \textit{Collector} and the transistor, between the \textit{Emitter} and the transistor, and between the \textit{Base} and the transistor. The meaning of all the transition rates are as follows: $W_{{Cj},x}^{(+X)}$ and $W_{{ij},x}^{(+X)}$ are the transition rates for the charge carriers $X$ ($X=P$ for holes and $X=N$ for electrons) from the cells (indexed by $Cj$ for \textit{Collector} cells and $ij$ for transistor cells) to their neighboring right cells in the $x$-direction; $W_{{Ej},x}^{(-X)}$ and $W_{{ij},x}^{(-X)}$ are the transition rates for the charge carriers $X$ from the cells (indices $Ej$ for \textit{Emitter} cells) to their neighboring left cells in the $x$-direction; $W_{{iB},y}^{(+X)}$ and $W_{{ij},y}^{(+X)}$ are the transition rates for the charge carriers $X$ from the cells (indices $iB$ for \textit{Base} cells) to their neighboring up cells in the $y$-direction; $W_{{ij},y}^{(-X)}$ are the transition rates for the charge carriers $X$ from the cells to their neighboring down cells in the $y$-direction; $W_{ij}^{(+)}$ and $W_{ij}^{(-)}$ are the transition rates for hole-electron pairs generation and recombination in transistor cells. The rates for transitions between cells are given by the product of the number of charge carriers $X$ in the departure cells and $D/\Delta x^2$ in the $x$-direction and $D/\Delta y^2$ in the $y$-direction except those transition rates across the junction. These exceptions are defined separately by
\begin{align}
& W_{{ij},x}^{(+X)}=\frac{DX_{i,j}}{\Delta x^2}\psi(\Delta U_{CB}^{(X)}) \hspace{0.3cm}\text{for}\hspace{0.2cm} i=\frac{l_C}{\Delta x} \text{,} \\
& W_{{ij},x}^{(-X)}=\frac{DX_{i,j}}{\Delta x^2}\psi(\Delta U_{BC}^{(X)}) \hspace{0.3cm}\text{for}\hspace{0.2cm} i=\frac{l_C}{\Delta x}+1 \text{,} \\
& W_{{ij},x}^{(+X)}=\frac{DX_{i,j}}{\Delta x^2}\psi(\Delta U_{BE}^{(X)}) \hspace{0.3cm}\text{for}\hspace{0.2cm} i=\frac{l_C+l_B}{\Delta x} \text{,} \\
& W_{{ij},x}^{(-X)}=\frac{DX_{i,j}}{\Delta x^2}\psi(\Delta U_{EB}^{(X)}) \hspace{0.3cm}\text{for}\hspace{0.2cm} i=\frac{l_C+l_B}{\Delta x}+1 \text{,}
\end{align}
where
\begin{align}
& \Delta U_{CB}^{(P)}=\Delta U_{BC}^{(N)}=e(\Phi_B-\Phi_C) \text{,} \\
& \Delta U_{CB}^{(N)}=\Delta U_{BC}^{(P)}=e(\Phi_C-\Phi_B) \text{,} \\
& \Delta U_{BE}^{(P)}=\Delta U_{EB}^{(N)}=e(\Phi_E-\Phi_B) \text{,} \\
& \Delta U_{BE}^{(N)}=\Delta U_{EB}^{(P)}=e(\Phi_B-\Phi_E) \text{,}
\end{align}
are the energy changes associated with the charge transitions and $\psi(\Delta U)$ is defined as
\begin{align}
\psi(\Delta U)\equiv\frac{\beta\Delta U}{\exp(\beta\Delta U)-1} \text{,}
\end{align}
guaranteeing the detailed balance condition in equilibrium,
\begin{align}
\psi(\Delta U)=\psi(-\Delta U)\exp(-\beta\Delta U) \text{.}
\end{align}
The transition rates for the hole-electron pairs generation and recombination are respectively given by
\begin{align}
W_{ij}^{(+)}=\Omega k_+ \hspace{0.3cm}\text{and}\hspace{0.3cm} W_{ij}^{(-)}=\Omega k_-\frac{P_{i,j}}{\Omega}\frac{N_{i,j}}{\Omega} \text{.}
\end{align}

\section{Stochastic Process of Langevin Type}\label{app_Langevin}

\par When the numbers of holes and electrons in each discretized cell are very large (typically greater than $100$), the master equation describing the stochastic evolution of the system state can be approximately expanded (Kramers-Moyal) up to the second order, giving the Fokker-Planck equation. This latter further leads to the stochastic process of Langevin type~\cite{Gaspard_NewJPhys_2005}. It is very intuitive to understand this Langevin stochastic process. After every small time step, the numbers of holes and electrons in each cell are updated due to the diffusion fluxes between the neighboring cells and the reaction flux in its own cell. These fluxes are expressed in terms of the transition rates in the master equation~(\ref{eq_master_equation}) and perturbed by Gaussian white noises. The number of holes $P_{i,j}$ obeys the stochastic differential equations of Langevin type,
\begin{align}
\frac{{\rm d}P_{i,j}}{{\rm d}t}=F_{i-1,j}^{(xP)}-F_{i,j}^{(xP)}+F_{i,j-1}^{(yP)}-F_{i,j}^{(yP)}+R_{i,j}+\delta_{iB}\delta_{j1}F_{i,B}^{(yP)} \text{,} \label{eq_dP_dt}
\end{align}
where $F_{i,j}^{(xP)}$ (respectively $F_{i,j}^{(yP)}$) is the stochastic flux of holes in the positive $x$-direction (respectively $y$-direction) at the $ij$-th cell, $R_{i,j}$ is the stochastic flux associated with the reaction in the $ij$-th cell, and $F_{i,B}^{(yP)}$ is the stochastic flux of holes in the positive $y$-direction between the \textit{Base} cell and transistor cell. For each $j$ in the range $1\le j\le M$, the $i$ appearing in $F_{i,j}^{(xP)}$ takes values from $0$ to $L$. Similarly, for each $i$ in the range $1\le i\le L$, the $j$ appearing in $F_{i,j}^{(yP)}$ takes values from $0$ to $M-1$. Those stochastic fluxes in Eq.~(\ref{eq_dP_dt}) with indices $i,\,j$ outside the above range are zero. The expressions of the stochastic fluxes are
\begin{align}
& F_{i,j}^{(xP)}=W_{i,j,x}^{(+P)}-W_{i+1,j,x}^{(-P)}+\sqrt{W_{i,j,x}^{(+P)}+W_{i+1,j,x}^{(-P)}}\xi_{i,j}^{(xP)}(t) \label{eq_dis_sto_1} \text{,} \\
& F_{i,j}^{(yP)}=W_{i,j,y}^{(+P)}-W_{i,j+1,y}^{(-P)}+\sqrt{W_{i,j,y}^{(+P)}+W_{i,j+1,y}^{(-P)}}\xi_{i,j}^{(yP)}(t) \label{eq_dis_sto_2} \text{,} \\
& R_{i,j}^{(P)}=W_{i,j}^{(+)}-W_{i,j}^{(-)}+\sqrt{W_{i,j}^{(+)}+W_{i,j}^{(-)}}\eta_{i,j}(t) \label{eq_dis_sto_3} \text{,} \\
& F_{i,B}^{(yP)}=W_{i,B,y}^{(+P)}-W_{i,1,y}^{(-P)}+\sqrt{W_{i,B,y}^{(+P)}+W_{i,1,y}^{(-P)}}\xi_{i,B}^{(yP)}(t) \text{,}
\end{align}
in terms of the Gaussian white noises $\xi_{i,j}^{(xP)}(t)$, $\xi_{i,j}^{(yP)}(t)$, $\eta_{i,j}(t)$, and $\xi_{i,B}^{(yP)}(t)$. The transition rate $W_{0,j,x}^{(+P)}$ is identified as $W_{Cj,x}^{(+P)}$ and $W_{L+1,j,x}^{(-P)}$ as $W_{Ej,x}^{(-P)}$. These Langevin stochastic equations are numerically simulated by discretizing time into equal intervals $\Delta t$ and replacing the white noises by independent and identically distributed Gaussian random variables. Similar expressions in this appendix hold for electrons. The stochastic differential equations~(\ref{eq_diff_1})-(\ref{eq_diff_4}) can be recovered in the continuum limit from Eqs.~(\ref{eq_dis_sto_1}) to (\ref{eq_dis_sto_3}) and similar equations for electrons. See Appendix~B in Ref.~\cite{Gu_PhysRevE_2018} for details.

\section{Dimensionless Quantities}\label{app_rescaling}

\par For numerical purpose, the values of physical quantities and parameters listed in Tables~\ref{tab_values_1} and~\ref{tab_values_2} are directly used in simulation. Accordingly, the numerical results are all given in terms of these values. However, all the quantities can be rendered dimensionless. This is achieved by introducing the characteristics quantities. The intrinsic carrier density $\nu=\sqrt{k_+/k_-}$ is used to define the dimensionless densities of all charge carriers
\begin{align}
n_*\equiv n/\nu \text{,}\hspace{0.5cm} p_*\equiv p/\nu \text{,}\hspace{0.5cm} a_*\equiv a/\nu \text{,}\hspace{0.5cm} d_*\equiv d/\nu \text{.}
\end{align}
The intrinsic carrier lifetime is introduced,
\begin{align}
\tau=\frac{1}{k_-\nu}=\frac{1}{\sqrt{k_+k_-}} \text{,}
\end{align}
so that we can define the dimensionless time
\begin{align}
t_*\equiv t/\tau \text{.}
\end{align}
The position is rescaled as
\begin{align}
x_*\equiv x/l_{\rm diff} \text{,}\hspace{1cm}\text{where}\hspace{1cm} l_{\rm diff}=\sqrt{D\tau}=\sqrt{\frac{D}{\sqrt{k_+k_-}}}
\end{align}
is the intrinsic carrier diffusion length before recombination. As a consequence of these definitions, the dimensionless current densities are given by
\begin{align}
j_{n*}\equiv\frac{j_n}{j_{\rm unit}} \text{,}\hspace{0.5cm} j_{p*}\equiv\frac{j_p}{j_{\rm unit}} \text{,}\hspace{1cm}\text{with}\hspace{1cm} j_{\rm unit}=\frac{\nu l_{\rm diff}}{\tau}=\sqrt{\sqrt{\frac{k_+}{k_-}}\cdot k_+D} \text{.}
\end{align}
The dimensionless electric field and potential are defined by
\begin{align}
{\cal E}_*\equiv\frac{{\cal E}}{{\cal E}_{\rm unit}} \hspace{1cm}\text{with}\hspace{1cm} {\cal E}_{\rm unit}=\frac{1}{l_{\rm diff}\beta e}=\sqrt{\frac{\sqrt{k_+k_-}}{D}}\cdot\frac{1}{\beta e} \text{,}
\end{align}
and
\begin{align}
\Phi_*\equiv\frac{\Phi}{\Phi_{\rm unit}} \hspace{1cm}\text{with}\hspace{1cm} \Phi_{\rm unit}=\frac{1}{\beta e} \text{.}
\end{align}
Using the values listed in Table~\ref{tab_values_1}, we have
\begin{align}
\nu=1 \text{,}\hspace{0.5cm} \tau=100 \text{,}\hspace{0.5cm} l_{\rm diff}=1 \text{,}\hspace{0.5cm} j_{\rm unit}=0.01 \text{,}\hspace{0.5cm}  {\cal E}_{\rm unit}=1 \text{,}\hspace{0.5cm} \Phi_{\rm unit}=1 \text{,}
\end{align}
and the relevant numerical results in the main text can be readily converted to the dimensionless ones. The dimensionless quantities can be compared with those obtained from experiments. In this way, numerical research can serve as a guiding role in devising electronic devices. Let us now consider a concrete example of a realistic silicon semiconductor whose typical values of the intrinsic density, intrinsic lifetime, and diffusion coefficient for charge carriers are~\footnote{Let us refer to the book by S. M. Sze and K. K. Ng, \textit{Physics of Semiconductor Devices}, Third Edition, Wiley, 2007. In Fig. 9 on page 20, we can find the the intrinsic density of silicon about $10^{10}{\rm cm}^{-3}$ at room temperature. In the Problem 33 and 34 on page 75, we can find the assumed diffusion coefficient $D=10\,{\rm cm}^2/{\rm s}$ and carrier lifetime $\tau=50\,\mu{\rm s}$, respectively. Since the parameter values are different from textbook to textbook, we only guarantee that the typical parameter values adopted in this work are correct on the orders of magnitude.}
\begin{align}
& \nu=1.5\times 10^{16}\,{\rm m}^{-3} \text{,} \\
& \tau=5\times 10^{-5} s \text{,} \\
& D=2\times 10^{-3} {\rm m}^2{\rm s}^{-1} \text{.}
\end{align}
The values of generation and recombination rate constants $k_+$ and $k_-$ can be calculated from equations $\nu=\sqrt{k_+/k_-}$ and $\tau=1/\sqrt{k_+k_-}$, reading
\begin{align}
& k_+=3\times 10^{20} {\rm m}^{-3}{\rm s}^{-1} \text{,} \\
& k_-=1.333\times 10^{-12} {\rm m}^3{\rm s}^{-1} \text{.}
\end{align}
The value of diffusion length $l_{\rm diff}=\sqrt{D\tau}$ is
\begin{align}
l_{\rm diff}=3.162\times 10^{-4} {\rm m} \text{.}
\end{align}
To fabricate a realistic bipolar $n$-$p$-$n$ junction transistor achieving the same (differential) amplification factor as that in the main text $\alpha\approx\bar{\alpha}\approx 164$, we should make sure the dimensionless sizes are the same. So, from Set III in Table~\ref{tab_values_2}, we have
\begin{align}
& l_C=l_E=m=l_{\rm diff}=3.162\times 10^{-4} {\rm m} \text{,} \\
& l_B=0.1l_{\rm diff}=3.162\times 10^{-5} {\rm m} \text{.}
\end{align}
In addition, the realistic transistor should have the same imbalances of densities of charge carriers at boundaries as those listed in Set III in Table~\ref{tab_values_2}. This requires that
\begin{align}
& \bar{n}_C=\bar{n}_E=10^{4}\nu=1.5\times 10^{20}\,{\rm m}^{-3} \text{,} \\
& \bar{p}_C=\bar{p}_E=10^{-4}\nu=1.5\times 10^{12}\,{\rm m}^{-3} \text{,} \\
& \bar{p}_B=10\nu=1.5\times 10^{17}\,{\rm m}^{-3} \text{,} \\
& \bar{n}_B=\nu/10=1.5\times 10^{15}\,{\rm m}^{-3} \text{.}
\end{align}
These boundary conditions can be realized by doping impurities properly. At the room temperature ($T=300 {\rm K}$), the value of the potential unit is
\begin{align} 
\Phi_{\rm unit}=\frac{1}{\beta e}=\frac{k_{\rm B}T}{e}\approx 0.0258 {\rm V} \text{,}
\end{align}
in terms of which the applied voltages can be quantified.

\printbibliography[title={References}]

\end{document}